\documentclass{aa}
\usepackage{graphicx}

\usepackage{times}

\begin{document}

\title{Ring shaped 6.7\,GHz methanol maser emission around a~young high-mass
star}

\author{A. Bartkiewicz \inst{1} 
        \and M. Szymczak \inst{1} 
        \and H.J. van Langevelde \inst{2,3} 
        }

\offprints{annan@astro.uni.torun.pl}
 
\institute{Toru\'n Centre for Astronomy, Nicolaus Copernicus 
          University, Gagarina 11, 87-100 Toru\'n, Poland 
\and      Joint Institute for VLBI in Europe, Postbus 2, 7990 AA 
          Dwingeloo, The Netherlands
\and      Sterrewacht Leiden, Postbus 9513, 2300 RA Leiden, The Netherlands
          } 

\date{Received 03 August 2005 / Accepted 14 September 2005}

\titlerunning{Ring shaped methanol maser emission}

\authorrunning{A.\,Bartkiewicz et al.}

\abstract{We report on EVN imaging of the 6.7\,GHz methanol maser
  emission from the candidate high-mass protostar G23.657-0.127. The
  masers originate in a nearly circular ring of 127\,mas radius and
  12\,mas width. The ring structure points at a central exciting object
  which characteristics are typical for a young massive star; its
  bolometric luminosity is estimated to be $\leq 3.2\times10^4{\rm
    L}_{\sun}$ and $\leq 1.2\times 10^5$L$_{\sun}$ for near
    (5.1\,kpc) and far (10.5\,kpc) kinematic distances,
    respectively. However, the spatial geometry of the
  underlying maser region remains ambiguous. We consider
    scenarios in which the methanol masers originate in a spherical
    bubble or in a rotating disc seen nearly face-on.
   }

\maketitle

\section{Introduction}
Methanol maser emission at 6.7\,GHz is a well established tracer of
high-mass star-forming regions (Menten\,\cite{menten91}). When studied
on milliarcsecond (mas) scales (a few hundreds of AU at the distances
of a few kpc) it shows various morphologies (Norris et
al.\,\cite{norris98}; Phillips et al.\,\cite{phillips98}; Walsh et
al.\,\cite{walsh98}; Minier et al.\,\cite{minier00}; Dodson et
al.\,\cite{dodson04}).
It has been argued that arc-like or curved structures can be produced 
by inclined discs (Norris et al.\,\cite{norris98}), while linear 
structures originate in a fraction of the discs which are seen exactly 
edge-on, resulting in strong masers through radial amplification 
(Minier et al.\,\cite{minier00}).

The hypothesis that 6.7\,GHz masers originate in circumstellar discs
has difficulty explaining approximately 60\% of the methanol masers,
which do not show a linear or curved morphology (Phillips et
al.\,\cite{phillips98}; Walsh et al.\,\cite{walsh98}).  It appears that
for most of those sources the maser morphology can be explained 
 by a shock wave model (Walsh et al.\,\cite{walsh98}).  
 Dodson et al. (\cite{dodson04}) suggested a model where
methanol masers form in planar shocks, and their velocity gradients
arise from the rotation of the underlying molecular cloud.
Although not in all cases a (detectable) H{\sc ii} region need to have
formed, such models are backed up by occasional close associations with
nearby (ultra-) compact H{\sc ii} regions. Elitzur (\cite{elitzur92})
argued that methanol masers arise in a layer of cool dense dust and gas
between the shock and ionization fronts around compact H{\sc ii}
regions. However, the lack of any observation of the expected
symmetric distribution of methanol emission was one argument against
this hypothesis.

In this letter we report on the discovery of a well-defined ring
structure in the 6.7\,GHz methanol maser line, coincident with the IR
detection of a young embedded star.  This distribution readily
  offers constraints on the origin of methanol masers by directly
  determining the separation of the excited region and the young star.
The striking geometry warrants a discussion of the underlying
three-dimensional structure of methanol masers.  We show that the
current observations can still be interpreted in more than one model,
but future observations will allow us to disentangle this geometry.

\section{Observations and data reduction}
The source G23.657$-$0.127 was detected in the unbiased Toru\'n survey
(Szymczak et al.\,\cite{szymczak02}) and displayed a rather complex and
relatively faint spectrum. Its position was determined subsequently
with a 0\farcs1 accuracy by MERLIN single-baseline observations. 
Then the methanol transition at 6668.519\,MHz was observed on 2004
November 11 with eight antennas (Cambridge, Darnhall, Effelsberg,
Medicina, Noto, Onsala, Toru\'n and Westerbork) of the European VLBI
Network (EVN){\footnote{The European VLBI Network is a 
  joint facility of European, Chinese, South African and other 
  astronomy institutes funded by their national research councils.}
as part of a larger sample. The total on-source time was
about 41\,min at different hour angles.  The phase-referencing scheme
used J1825$-$0737 (240mJy at 6.7\,GHz), separated by 2\fdg4 from the
target.  The bandwidth was 2\,MHz in both circular hands, covering 
  LSR velocities from 52 to 141\,km\,s$^{-1}$, divided into
1024 channels yielding a velocity resolution of 0.09\,km\,s$^{-1}$.

The data calibration and reduction were carried out with standard
procedures for spectral line observations using the Astronomical Image
Processing System (AIPS).  The phase referencing yields absolute
  position of the target and the accuracy is estimated to be 12\,mas
in Dec and 10\,mas in RA.  For detailed analysis the target was then
self-calibrated on a strong (3.6\,Jy) and point-like maser spot
identified at 82.6\,km\,s$^{-1}$.  An area of 1$\times$1\,arcsec$^2$
was searched for emission over the entire band. The analysis was
  carried out on images obtained with natural weighting and a
  resulting beam of 5.5\,mas$\times$16\,mas at a position angle of
$-1\degr$. The resulting rms noise level (1$\sigma$) in line-free
channels was 3.7\,mJy\,beam$^{-1}$. These observations are among the
first EVN 5\,cm results with 8 antennas, and the superior image quality
allows the detection of many weak features. The calibration and
  data reduction procedures will be described in more detail in a
forthcoming paper.

\section{Results}

\begin{figure}
   \includegraphics[height=20cm]{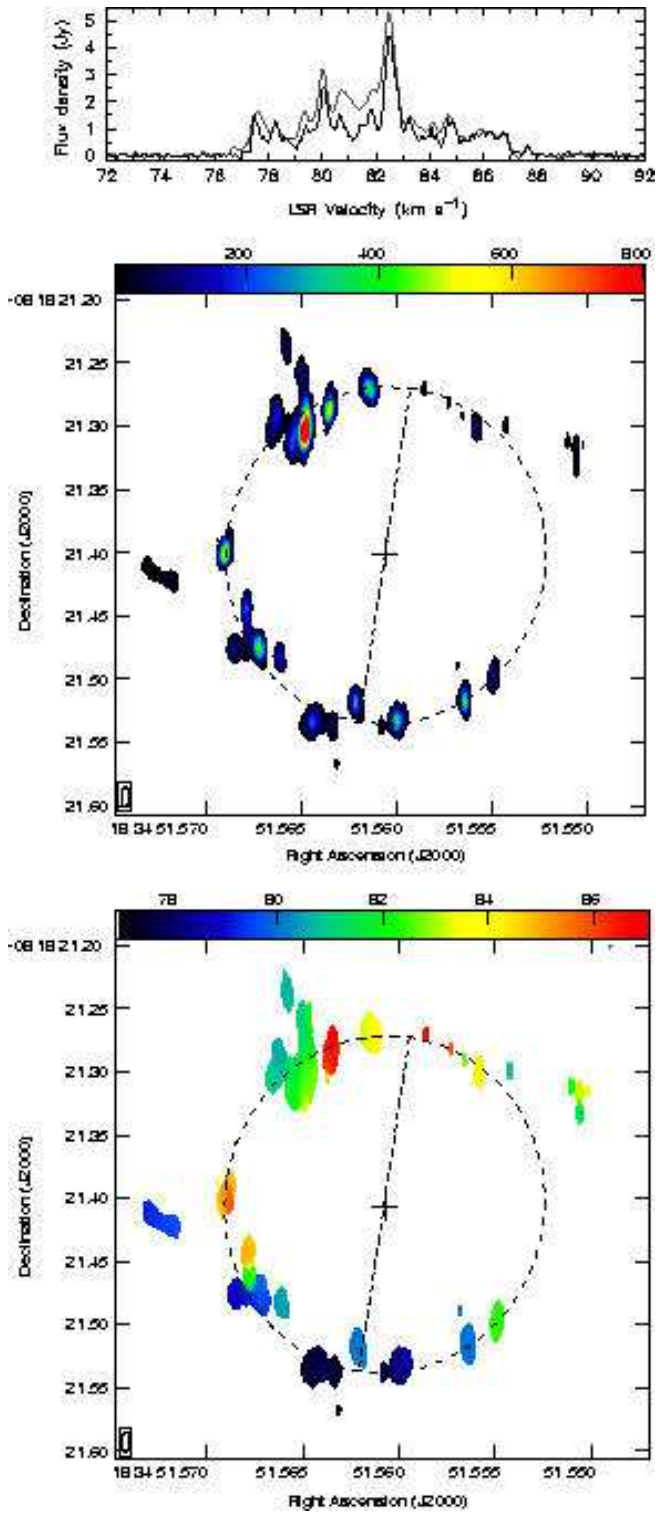}
   \caption{The 6.7\,GHz methanol maser from G23.657$-$0.127. {\it Top}: Spectrum
     of the integrated flux density from all the maser emission in the
     map (bold line) together with the Toru\'n total power spectrum
     (thin line) taken on 2005 February 2.  {\it Middle}: Total
     intensity (zeroth moment) map. The colour scale varies linearly
     from 5 to 800\,Jy\,beam$^{-1}$\,m\,s$^{-1}$. The beam is indicated by
     the ellipse in the bottom left-hand corner of the image. The
     dashed ellipse shows the flux-weighted fit to the data while 
     the dashed line indicates the orientation of the major axis.
     The cross indicates the inferred position of 
     a central object.  {\it Bottom}: Velocity field of maser components 
     (first moment map). The colour scale varies linearly from 77.0 to
     87.0\,km\,s$^{-1}$.}
   \label{maps}
\end{figure}

Methanol maser emission was detected over a range of 10.8\,km\,s$^{-1}$
between 77.0 to 87.8\,km\,s$^{-1}$ (Fig. \ref{maps}). This velocity
range is similar to that reported for other sources (Szymczak et
al.\,\cite{szymczak05}).  The central velocity of the methanol maser
profile is 82.4\,km\,s$^{-1}$.  An image integrated over all spectral
channels containing emission (Fig. \ref{maps}) shows that the masers
are distributed in a ring.  We registered 315 maser spots brighter than
10$\sigma$ in individual channel maps, which were found to be
clustered into 31 maser components. Their brightness temperatures range
from 2$\times$10$^7$ to 10$^9$\,K. The masers appear to be absent from
the western portion of the ring and only weak emission is seen in the
north-west. We note that about 31\% of the flux is missing in
the VLBI data, as compared to the single dish observation
(Fig.\ref{maps}). Taking into account a 10\% accuracy of the flux
calibration of both observations, there is a significant excess of the
single dish emission over the velocity range 80.6 to
82.2\,km\,s$^{-1}$, resolved out by the VLBI observations.

Rigorous analysis of the data shows that the distribution of maser
components can be fitted by an ellipse with major and minor semi-axes
of 133 and 123\,mas, respectively. The major axis is 
elongated along the position angle of $-$9\fdg6 (Fig. \ref{maps}). 
The position of the centre (Table \ref{pos}) was determined by
minimizing the distance between all the maser spots (flux-weighted)
and the model ellipse.

Since the maser morphology differs by less than 8\% from a circular
ring we used the AIPS task IRING to determine the radial distribution
of the maser emission (Fig. \ref{radprof}). The mean radius is
127\,mas and the width that contains 50\% of the flux is 12\,mas.  The
inner and outer radii measured at 10\% and 90\% of the cumulative flux
are 116 and 145\,mas, respectively.  We conclude that the methanol
masers originate in a thin shell with a width of 29\,mas ($\sim$20\% of
the radius), which has a sharp inner edge and a more shallow outer
border.

\begin{table*}
\caption {The position of the centre of the G23.657$-$0.127 
maser and the infrared counterparts.}
\begin{tabular}{l l l l l}
\hline
           & RA(J2000)(\fs)      & Dec.(J2000)(\farcs) & Difference from  & Reference \\
           & (18$^{\rm h}$34$^{\rm m}$)& ($-$08\degr18\arcmin)& radio position (\farcs) &            \\
\hline
G23.657$-$0.127 & 51.5606$\pm$0.0007 & 21.401$\pm$0.012 &  &  this paper \\
2MASS183451.56$-$0818214 & 51.56$\pm$0.0073(3)   & 21.4$\pm$0.11 & 0.087 & 2MASS (Cutri et al.\cite{cutri03}) \\
G023.6566-00.1273 & 51.6$\pm$0.02         & 22$\pm$0.3 & 0.6 & MSX6C (Egan et al.\cite{egan03}) \\
IRAS18321$-$0820  & 52.0$\pm$0.96         & 20$\pm$9.5 & 6.1 & IRAS PSC (IPAC \cite{ipac86}) \\ 
\hline
\end{tabular}
\label{pos}
\end{table*}

\begin{figure}
   \includegraphics[height=5.2cm]{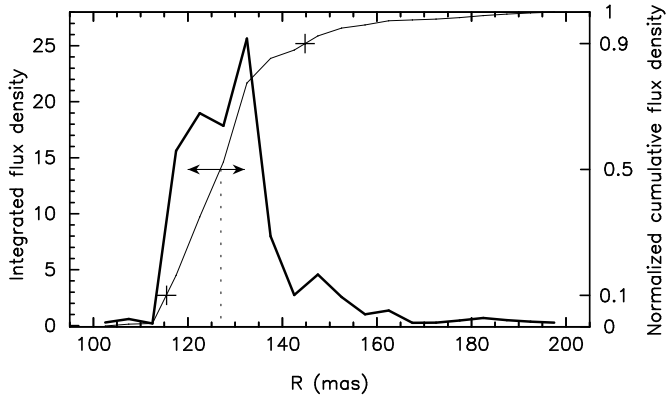}
   \caption{The integrated flux density per annulus of 5\,mas (bold line)
     and the normalized cumulative flux density (thin line) versus
     radius. The radius of the maser ring at 50\% of
     the total emission is marked by the dotted line.  The width
     between a normalized cumulative flux density range of 25\% to 75\%
     (and 10\% and 90\%) is marked by the arrowed horizontal bar (and
     crosses).}
   \label{radprof}
\end{figure}

Looking at the detailed velocity structure of the masers, we found that
15 out of 31 maser components display clear spatially coherent
filaments or arcs. They have sizes from 4\,mas to 29\,mas 
  with internal velocity gradients from 16\,m\,s$^{-1}$\,mas$^{-1}$ to
168\,m\,s$^{-1}$\,mas$^{-1}$.  Fig. \ref{v-azimuth} shows the velocity
of all components versus the azimuth angle between the maser spot and
the major axis of the ellipse fitted in Fig. \ref{maps}. Considerable
velocity dispersion of $\sim$5.4\,km\,s$^{-1}$ exists, but there
  is also a weak signature detectable with the dominant blue- and
red-shifted emission originating from the southern and northern parts
of the ring, respectively. It is also remarkably that the velocity
gradients of the masers components are all dominantly radial.

\section{Discussion}

The most striking result from our VLBI observations is the detection of
a nearly circular ring of maser emission.  The symmetric
distribution of maser components strongly suggests that they have a common
origin and that there exists a central source. It is very remarkable and
reassuring that the ring centre coincides with the infrared source
2MASS183451.56$-$0818214 within the position uncertainties.  It also
matches (2$\sigma$ position uncertainty) with the object
G023.6566-00.1273 in the MSX6C catalogue and lies well within the
position error ellipse of IRAS18321$-$0820 (Table~\ref{pos}). Although
one must remember that the infrared data were taken with a much coarser
resolution, we derived the spectral energy distribution of the object,
which is typical for embedded protostar(s) or recently formed high-mass
star(s) (Fig.~\ref{sed}). A model of two black-body components 
(Walsh et al.\,\cite{walsh99}) reproduces the data satisfactory; the 
cold dust temperature is 80\,K and the hot dust temperature is 540\,K.
We note that this is a very crude estimate, as the 60 and 100\,$\mu$m
flux densities are poorly determined and millimeter wavelength emission
was not yet measured. No radio continuum emission was found above a
detection limit of $\sim$3\,mJy at 5\,GHz (Giveon et
al.\,\cite{giveon05}). Searches for other masers species were negative: 
neither H$_2$O masers at 22\,GHz, nor any of the four OH masers at
1.6\,GHz were found (Szymczak \& G\'erard \cite{szymczak04}; 
Szymczak et al.\cite{szymczak05}).
Only weak absorption features near 80\,km\,s$^{-1}$
appeared at 1665 and 1667\,MHz (Szymczak \& G\'erard
\cite{szymczak04}).

\begin{figure}
   \includegraphics[height=5.3cm]{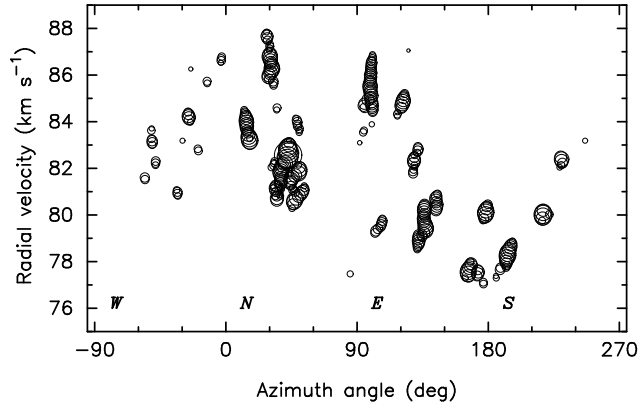}
   \caption{Velocity of the maser spots in the ring versus azimuth angle 
     measured from the major axis (north to east). The sizes of circles are
     proportional to the logarithm of the flux densities.}
   \label{v-azimuth}
\end{figure}

Assuming that the central velocity of the methanol maser profile
(82.4\,km\,s$^{-1}$) is the systemic velocity and using the equation
for the Galactic rotation curve given by Brand \& Blitz
(\cite{brand93}), we derived the near and far kinematic distances of
G23.657$-$0.127 as 5.1 and 10.5\,kpc, respectively. Applying the 
formula by Walsh et al. (\cite{walsh97}; their equation 3) we estimated the 
bolometric luminosities from the mid- and far-infrared emission 
as: $\leq 3.2\times10^4$L$_{\sun}$ and $\leq 1.2\times
10^5$L$_{\sun}$ for the near and far kinematic distances, respectively.
If these luminosities are provided by single stars then these 
would correspond to B0 and O7 ZAMS stars (Panagia \cite{panagia73}). The 
luminosities are, however, most likely due to a cluster of 
stars, unresolved in the infrared (Walsh et al.\,\cite{walsh97}, \cite{walsh01}).
For the near and far kinematic distances the average 
radius of the maser ring as determined in Sect.~3 is 650\,AU 
and 1330\,AU and its width is 60\,AU and 130\,AU, respectively.

In principle there are several three-dimensional structures that
project onto a ring structure like we have observed. The fact that the
maser components at velocities close to the central velocity resolve
out into multiple directions can be interpreted as a circumstellar
shell or a spherical bubble. However, a disc geometry seen face--on
results in a similar structure. We discuss these options below.

A steep increase of the maser intensity at the inner edge of the ring
and a smooth decrease at its outer edge suggest that the maser arises
in a narrow circular layer of the excited material. One can imagine
that the masers outline an expanding bubble and we observe a ring from
the material in plane of the sky through tangential amplification. The
bubble may be the shock front originating from the central star and
propagating into the circumstellar gas. Assuming an expansion
  velocity of 5.4\,km\,s$^{-1}$ its dynamical age is $\sim$550\,yr or
  $\sim$1130\,yr for the near and far kinematic distances, respectively.  
A spherical bubble was recently also detected in Cep\,A by H$_2$O
maser emission (Torrelles et al.\,\cite{torrelles03}) and  may
  occur as a short lived stage in the earliest stages of stellar
evolution.  The lack of detectable radio continuum at 5\,GHz does not
preclude the interpretation as a spherical shock from a young
  stellar object. It is quite possible that the ionization front
around the central star(s) of G23.657$-$0.127 is too weak to be visible
at 5\,GHz or so dense that it becomes detectable only at higher
frequencies (Carral et al.\,\cite{carral96}). Alternatively, the
  result of shock wave may be limited to dense knots of compressed and
  accelerated gas without ionizing it (Phillips et al.\,\cite{phillips98}).
The radial shock wave model offers a natural explanation of the 
   observed filaments and arcs with internal velocity gradients. 

The observed weakly elliptical structure (0.38 eccentricity) can be
interpreted as a disc inclined at an angle of 68\degr. However,
  for a geometry so close to face-on the inclination and the
  orientation of the major axis are poorly constrained. One expects a
  weak velocity signature with the extreme velocities originating from
  where the ellipse intersects with the major axis. Such a velocity
  signature between the south and the north may be detected (Sect. 3),
  but the large uncertainty does not allow a convincing fit of a
  rotating disk to the data (Fig. \ref{v-azimuth}). 
 A~complication of this model is that the methanol masers would need to
  be built up nearly perpendicular to the disc plane.

\begin{figure}
   \includegraphics[height=5.4cm]{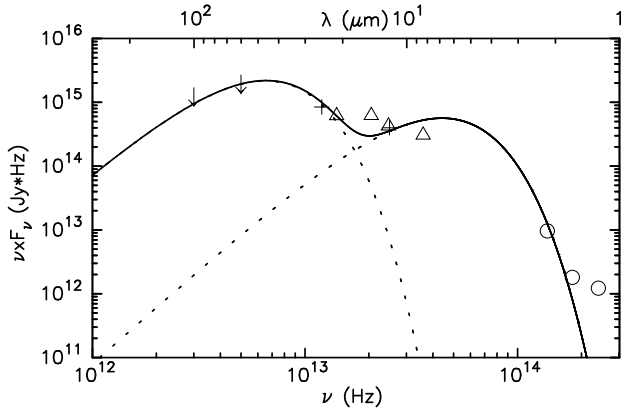}
   \caption{Spectral energy distribution of G23.657$-$0.127. Data for
     counterparts (Table 1) in 2MASS
    (circles), MSX6C (triangles) and IRAS (crosses)  
    are shown. The upper limits for IRAS are marked by arrows. A model of
    two black-body components (Walsh et al.\,\cite{walsh99}) 
    is represented by the curved lines.}
   \label{sed}
\end{figure}

Previous claims of methanol discs all have derived edge-on geometries.
This can be understood as a selection effect, because the strongest
masers may result from radial amplification. The current observations
focused on sources with relatively low peak fluxes. We have been able
to image many weak features because of the increased number of EVN 
antennas with 5\,cm receivers. 
With the currently available data it is not possible to distinguish between
the spherical bubble and the rotating disc models. Further observational 
verifications of these possibilities include high resolution studies of
radio continuum, tracers of shock fronts as well as multi-epoch studies of
methanol masers.

\begin{acknowledgements}
  We thank to Bob Campbell at JIVE for his detailed support in many
  stages of this experiment and to an anonymous referee for
  useful comments. This work has benefited from research
  funding from the EC 6th Framework Programme and supported by the MNII 
  grant 1P03D02729. 
\end{acknowledgements}


\begin{thebibliography}{}
\bibitem[1993]{brand93} Brand, J., \& Blitz, L. 1993, A\&A, 275, 67
\bibitem[1996]{carral96} Carral, P., Kurtz, S.E., Rodriguez, L.F., de Pree, C., \& Hofner, P. 1996, ApJ, 486, L103 
\bibitem[2003]{cutri03} Cutri, R.M., Skrutskie, M.F., Van Dyk, S., et al. 2003, 
        2MASS All-Sky Catalog of Point Sources,
        ftp://cdsarc.u-strasbg.fr/pub/cats/II/246 
\bibitem[2004]{dodson04} Dodson, R., Ojha, R., \& Ellingsen, S.P. 2004, MNRAS, 351, 779
\bibitem[2003]{egan03} Egan, M.P., Price S.D., Kraemer, K.E., et al. 2003,
        MSX6C Infrared Point Source Catalog,
        ftp://cdsarc.u-strasbg.fr/pub/cats/V/114
\bibitem[1992]{elitzur92} Elitzur, M. 1992, Astronomical Masers, Kluwer, Dordrecht
\bibitem[2005]{giveon05} Giveon, U., Becker, R.H., Helfand, D.J., \& White, R.L. 2005, AJ, 129, 348
\bibitem[1986]{ipac86} Joint IRAS Science W.G. 1986,
        IRAS Catalog of Point Sources, Version 2.0,
        ftp://cdsarc.u-strasbg.fr/pub/cats/II/125
\bibitem[1991]{menten91} Menten, K.M. 1991, ApJ, 380, L75
\bibitem[2000]{minier00} Minier, V., Booth, R.S., \& Conway, J.E. 2000, A\&A, 362, 1093
\bibitem[1998]{norris98} Norris, R.P., Byleveld, S.E., Diamond, P.J., et al. 1998, ApJ, 508, 275
\bibitem[1973]{panagia73} Panagia, N. 1973, AJ, 78, 929
\bibitem[1998]{phillips98} Phillips, C.J., Norris, R.P., Ellingsen, S. P., \& McCulloch, P. M. 1998, MNRAS, 300, 1131
\bibitem[2002]{szymczak02} Szymczak, M., Kus, A.J., Hrynek, G., Kepa, A., \& Pazderski, E. 2002, A\&A, 392, 277 
\bibitem[2004]{szymczak04} Szymczak, M., \& G\'erard, E. 2004, A\&A, 423, 209
\bibitem[2005]{szymczak05} Szymczak, M., Pillai, T., \& Menten, K.M. 2005, A\&A, 434, 613
\bibitem[2003]{torrelles03} Torrelles, J.M., Patel, N.A., Anglada, G., et al. 2003, ApJ, 598, L115
\bibitem[2001]{walsh01} Walsh, A.J., Bertoldi, F., Burton, M.G., \& Nikola, T. 2001, MNRAS, 326, 36
\bibitem[1998]{walsh98} Walsh, A.J., Burton, M.G., Hyland, A.R., \& Robinson, G. 1998, MNRAS, 301, 640
\bibitem[1999]{walsh99} --- 1999, MNRAS, 309, 905
\bibitem[1997]{walsh97} Walsh, A.J., Hyland, A.R., Robinson, G., \& Burton, M.G. 1997, MNRAS, 291, 261
\end{thebibliography}
\end{document}